\documentstyle[12pt]{article}
\renewcommand{\baselinestretch}{1.47}
\def\eps{\epsilon }

\def\part{\partial } 
 
\def\Yv{\vec{Y} }                                                      
                                                      
\def\alv{\vec{\alpha} }                                                      
\def\alvh{\hat{\vec{\alpha}} }                                                      
\def\gv{\vec{g} }                                                      
\def\tv{\vec{t} }                                                      
\def\yv{\vec{y} }                                                      
\def\muv{\vec{\mu} }                                                      
                                                      
\def\tb{\overline{t} }

\def\Ybf{\bf Y\,}

\def\albf{{\bf \alpha}\, }
\def\alh{\hat{\bf \alpha}\, }

\def\Sibf{{\bf \Sigma}\, }

\def\Abf{{\bf A}\,}
\def\Abfh{\hat{\bf A}\,}
\def\Dbf{{\bf D}\,}
\def\Hbf{{\bf H}\,}

\def\Mbf{{\bf M}\,}
\def\Obf{{\bf O}\,}

\def\Ubf{{\bf U}\,}

\def\Labf{{\bf \Lambda}}

\def\uv{\vec{u}}

\def\sgh{{\hat \sigma}}

\def\btvh{\hat{\vec{ \beta}}}
\def\btv{\vec{ \beta}}
\def\lahv{\vec{\hat{\lambda}}}
\def\lav{\vec{ \lambda}}

\def\np{\vfill\eject}       
\def\ni{\noindent}
\def\IR{I\kern-.255em R}

\def\eps{\epsilon }

\def\Xbf{{\bf X}\,}
\def\Sibf{{\bf \Sigma}\, }
\def\Sibfh{{\bf {\hat\Sigma}}\, }

\def\Sbf{{\bf S}\, }

\def\Rh{\hat{R}\, }

\def\eps{ \epsilon}

\def\btv{\vec \beta}

\def\Ab{{\bf A}\,}
\def\Abf{{\bf A}\,}

\def\Cbf{{\bf C}\,}
\def\Ebf{{\bf E}\,}
\def\Gbf{{\bf G}\,}

\def\Mbf{{\bf M}\,}

\def\Qbf{{\bf Q}\,}
\def\Sbf{{\bf S}\,}
\def\Labf{{\bf \Lambda}}

\begin{document}
\begin{center}
{\bf SMOOTHING SPLINE GROWTH CURVES \\
WITH COVARIATES}\\
\ \\
{\bf Kurt S. Riedel} \ \ 
and \ \ {\bf Kaya Imre}$^*$ \\ 
Courant Institute of Mathematical Sciences \\
New York University \\
251 Mercer St. \\
New York, NY 10012\\
\ \\
\end{center}

\ni

\ \\
\centerline{ABSTRACT}

We adapt the interactive spline model of Wahba to growth curves 
with covariates. The smoothing spline formulation permits a nonparametric
representation of the growth curves.
In the limit when the discretization error is small relative to the 
estimation error,
the  resulting growth curve estimates often depend only weakly 
on the number and locations of the knots. 
The smoothness parameter is determined from the data by minimizing
an empirical estimate of the expected error. We show that the risk
estimate of Craven and Wahba is a weighted goodness of fit estimate.
A modified loss estimate is given, where $\sigma^2$ is replaced by its
unbiased estimate.  
\ \\
\ \\
$^*${Permanent address: College of Staten Island, C.U.N.Y., Staten Island}

\np
\ni
{\bf I.  INTRODUCTION}
\ \\	
Growth curve analysis is used to parameterize a family of temporal
curves whose shapes depend on a vector of covariates $\vec{u}$
(Potthoff and Roy (1964),
Grizzle and Allen (1969), Geisser (1980)).
As an example, we consider the heights of children as a
function of time or age and of a number of covariates; both discrete
covariates such as sex, and continuous covariates such as drug
dosages. We are primarily interested in the case where each individual
has been measured a large number of times, so that the growth as 
a function of time may be treated nonparametrically. 
We assume that the number of individuals is small enough, or the covariate
dependencies are simple enough, that the covariate dependencies may be
treated parametrically for each fixed time.

The general theory of growth curves allows arbitrary basis functions
and Bayesian priors (Lee and Geisser (1972), 
Rao (1975), Strenio et al. (1983)). In practice, however, the
basis functions are usually assumed to be polynomials in time.
Similarly, the covariance matrix of the Bayesian prior is 
usually a diagonal matrix, or
is determined empirically from repeated measurements, or is given
a low order autoregressive structure. 
Smoothing splines is a powerful technique 
used to reconstruct nonparametric curves and surfaces
(see Silverman (1985), Eubank (1988), Muller (1988), Wahba (1990) for reviews). 
The smoothness penalty function corresponds to a
Bayesian prior. Recently, Wahba has developed a general theory of interactive
smoothing spline models (Ch. 10, Wahba (1990)). In this brief article, 
{\it we
adapt interactive spline models to  growth curve analysis.}
Our nonparametric growth curve models may also be viewed as a variant
of the $\Pi$ model of Breiman (1991)
which is appropriate when only one of
the covariates is ``timelike'' and therefore needs to be treated
nonparametrically. 

Smoothing techniques are widely used for the nonparametric determination
of a single growth curve, including the case of repeated
measurements (Gasser et al. (1984)). For families
of covariate growth curves, however, there has been surprisingly little
nonparametric work. 
Partial linear models (Rice (1986), Heckman (1986), Speckman (1988)) consider
growth curves where the parametric part of the model depends
on covariates, but the nonparametric part depends only on time: 
$\mu(t,\vec{\xi}_i,i)=\vec{\xi}_i \cdot \btv + f(t)$.
Raz (1989) considers families of grouped growth curves, and uses smoothing
splines to determine both the individual growth curves and the main
effect growth curves. Due to the repeated measurement structure of his
model, he is able to determine each nonparametric function separately.
A multidimensional nonparametric ANOVA decomposition is given in
Gu and Wahba (1991).

Of the two main smoothing techniques, kernel smoothing and smoothing
splines,  we concentrate on smoothing splines because they generalize
more naturally to growth curves with covariates.
Despite its generality, we are unaware of any growth curve analysis where
the nonparametric part of the model depends on covariates through the
standard sum of products structure.
Similarly, except our own research
(Kardaun et al. (1988), McCarthy et al. (1991), Riedel (1992)),  
we are unaware of any growth curve
analysis with the sum of products covariate structure 
where the basis representations are regression splines. 

We briefly review smoothing splines for a single growth curve
with $n$ independent measurements: $y_i = f_o(t) +\eps_i$ 
where $\eps_i \sim N(0,\sigma^2)$. We represent the curve as a sum
of cubic spline basis functions:
$f_o(t) = \sum_{k_{} =1}^{K} \alpha_{k} B_k (t)$. In the cubic spline model,
the model is a piecewise cubic polynomial with a discontinuous third 
derivative at each knot locations. When the knots are in the interior 
of the domain, the number of free parameters
is equal to the number of knots plus four. 
The free parameters, $\alpha_k$, are determined by 
minimizing the functional
$$
\min_{\alv} \sum_i^n {( {y}_i - f_o(t_i,\alv))^2 \over \sigma^2}
+
\lambda_{o} \int_0^1 (f_{o}^{(\gamma)} (t,\alv))^2 dt \ ,
\eqno (1)$$
where $\gamma$ is a positive integer with $2 \le \gamma \le 3$.
When the smoothing parameter, $\lambda_o$, is zero or
too small, the spline coefficients become ill-conditioned and spurious
oscillations develop with a wavelength proportional to twice the the distance
between knots (Wegman and Wright (1983)). 
Two remedies for this ill-conditioning are to decrease the number of basis
functions and to increase the smoothness parameter $\lambda_o$.
Adjusting the spline representation is usually a very complicated optimization.
Therefore researchers have concentrated on finding methods which determine
the optimal value of $\lambda_o$ from the data.
The smoothness penalty function then shrinks the {\it a priori} probability of
oscillatory behavior to a small and acceptable level.

For a single function, 
the minimization of Eq. 1 can be done over an infinite dimensional
Sobolev space, and the resulting minimum is a cubic spline function with
knots at all the measurement locations.   
In practice, many fewer knots 
may be sufficient to represent the unknown function, $f_o(t)$. This
class of representations is called hybrid spline models.
In the covariate growth curve models of Sec. II, 
the measurement times are usually nonuniform, and we normally consider hybrid
spline models with significantly fewer knots than the total number of distinct
measurement times. 

When hybrid splines are used, two types of error arise: model error from
discretization effects and estimation error.  Agarwal and Studden (1980)
give bounds on the discretization error. We present generalizations of 
the standard estimation error formulas in Sec. III. Detailed asymptotic 
analysis can be found in  Cox (1984), Eubank (1988),  and Wahba (1990).
When discretization error is small relative to the estimation error,
hybrid models are generally appropriate since they reduce the ill-conditioning
of the estimate at relatively little cost in terms of the total error.
For the profile data of Sec. V, we find that
the fitted function depends only weakly on the locations of the knots. 

\ \\
\ni
{\bf II.  SEMIPARAMETRIC GROWTH CURVE MODELS}
\bigskip

For covariate growth curves,
we consider a family of $N$ individuals with the $i$th individual
having measured responses at $n_i$ timepoints, 
$t_{1,i}, \ldots t_{n_i,i}$. 
We assume the covariates are fixed in time for the $i$th individual and
denote their value by the $M$ vector, $\uv_i$.
We denote the $n_i$ vector of the $i$th measurement times,
$(t_{1,i}, \ldots t_{n_i,i})^t$, by $\tv_i$. 
We denote the $n_i$ vector of measured responses,
$(y_{1,i}, \ldots y_{n_i,i})^t$, by $\yv_i$ and the vector of true values,
$(\mu(t_{1,i},\uv_i,i), \ldots \mu(t_{n_i,i},\uv_i,i))^t$, 
by $\muv_i(\tv_i,\uv_i)$. 
Although we are primarily interested in data which is grouped by individual,
multiple time measurements of each individual are not necessary.
Uncorrelated measurements may be treated with $n_i =1$.

 We divide the model for 
$\mu(t,\vec{u},i)$ into a parametric part, $h(t,\vec{u},i)$, 
and a nonparametric part, $f(t,\vec{u},i)$.  
The parametric part of the model 
is assumed to have a specific known functional form:
 $h(t,\uv,\btv,i)  = \sum_{j_{} =1}^{J} h_j(t,\uv,i) \beta_j$, 
where the $h_j(t,\uv,i)$ are known functions and the $\beta_j$
are undetermined free parameters. We allow both the parametric 
and nonparametric terms to
depend on the specific individual to allow for fixed effects models.

We therefore consider growth curve models of the form
$$
\mu(t,\vec{u},i) = h(t,\uv,\btv,i) +
f_0 (t) + \sum_{\ell =1}^L f_{\ell} (t)
g_{\ell} ( \vec{u},i) \ .\eqno (2a)
$$
We assume that the temporal structure of the $f_{\ell}(t)$ is sufficiently
complex that it requires a nonparametric representation. 
The model of Eq. 2a is a special case of the $\Pi$ model of Breiman (1991). 
We assume that the covariate dependencies, $g_{\ell}(\uv,i)$, are either
given or that they depend on a low  number
of free parameters, $\btv$: $g_{\ell} ( \vec{u},i,\btv)$.
In contrast, Breiman estimates
both $f_{\ell}(t)$ and $g_{\ell}(\uv,i)$ nonparametrically.

When the last term, $\sum_{\ell =1}^L f_{\ell} (t)
g_{\ell} ( \vec{u},i)$, is omitted, our models reduce to the standard
partial linear spline models.
Our sum of products covariate spline models resemble generalized
additive models, and we expect many of the same techniques and theorems
to apply (Friedman and Stuetzle (1981), Hastie and Tibshirani (1990),
Friedman (1991)). 
Similar to  Wahba's interactive spline model, we use a separate smoothing
spline for each of the $f_{\ell} (t)$.

In most applications, the covariate basis functions,
$g_{\ell} ( \vec{u} )$, are linear, $(u_m - \bar{u}_m$), or quadratic, and are
centered about the database mean or are discrete variables.
If the nonparametric part of each individual curve is considered noise,
a nonparametric function may be included for each separate individual:
$g_{\ell}(\uv,i)= \delta_{\ell,i}$. 

In our initial study of profile variation 
(Kardaun et al. (1988), McCarthy et al. (1991)),
the data consisted of $N$ sets of temperature measurements at $15$
different radial locations in a fusion plasma, with $10 \le N \le 100$.
The radial location, $r$, is the timelike variable, and the main covariate
is $q$, which corresponds to the ratio of the average magnetic field to 
the total plasma current. 
The shape of the profiles is much less variable than the overall
magnitude; therefore each individual curve was given its own
own intercept: $h_j(t,\uv,i)= \delta_{j,i}$. Thus the log linear nonparametric
model of McCarthy et al. (1991) is
$$
ln[T(r,q,i)] = \beta_i + f_0 (r) + f_{1} (r)ln(q) 
\ . \eqno (2b)$$
In our earlier study, we used regression splines ($\lambda \equiv 0$) with
only three knots. In Section V, we consider data with more than 60
measurements per profile. This data requires many more knots, and the 
resulting estimation problem is ill-conditioned without a smoothness
penalty function. 

We assume that both the number and choice of both 
the parametric basis functions, $h_j(t,\uv,i)$, and the covariate basis 
functions, $g_{\ell}(\uv,i)$ are given {\it a priori}. In practice, both sets
of basis functions are often determined iteratively using a combination
of statistical common sense and the  
minimizations of  weighted sums of residual errors and smoothness/degree of 
freedom/predictive error penalty functions.

Each of the functions,      
$f_{\ell} (t)$, is given a radial representation:
$f_{\ell} (t) = \sum_{k_{} =1}^{K} \albf_{k \ell} B_k (t),$
where the $B_k (t)$ are basis functions. We recommend the choice of $B$
splines for the $B_k$ (deBoor (1978)).
$B$ splines are a particular reparameterization
of the standard piecewise polynomial splines 
which have the advantage that each function
is spatially localized. The resulting design matrix has a band
structure. In practice, the innermost and outermost basis functions are 
often restricted to be linear in time.

When the temporal design is uniform, $\tv_i = \tv_j$, the 
knot positions may be chosen at all the distinct measurement times.
If each individual is measured at slightly different times,
$t_{p,i} = \tb_p + \tilde{t}_{p,i}$, we can choose the knot locations at
the typical measurement times, $\tb_p$. Alternatively, we can place a
knot at every distinct measurement time: $t_{p,i}$. The value of the additional
knots in reducing the model error depends on the ratio of the 
typical time between measurements, $\tb_{p+1}-\tb_p$, to the 
the spread of the measurement times, $\sigma_{tp}$ where
$\sigma_{tp}^2 \equiv {1\over N}\sum_{i=1}^N \tilde{t}_{p,i}^2$.
In many cases, the spread of times is small relative to the 
typical time between measurements.
In these cases,
we typically choose the first alternative, knot locations only
at $\tb_p$. Alternatively,
the number and location of the knot positions can be determined 
by convergence tests or by the data-based parameter selection criteria of 
Sec. III.  

We denote the $K \times (L+1)$ matrix whose elements are
$\albf_{k, \ell}, \ell=0,\ldots L,\ k=1,\ldots K$ by $\albf$, 
and the $\ell$th column
vector by $\alv_{\ell}$: $\alv_{\ell_k}\equiv \albf_{k, \ell}$. 
The $n_i \times K$
temporal design matrix for the nonparametric part of the $i$th 
individual is denoted
by $\Xbf_i$ with elements $X^i_{p,k} = B_k(t_{p,i})$.
Similarly, the parametric part of the design matrix is defined by
$H^i_{p,j'} = h_{j'}(t_{p,i},\uv_i,i)$.
Thus the  $i$th individual is parameterized by
$$
\muv_i(\tv_i,\uv_i) = \Hbf_i \btv +\Xbf_i \albf \vec{g} ( \vec{u}_i,i)
\ , \eqno (3)$$
where $\gv(\uv_i,i) \equiv (1, g_1(\uv_i,i),\ldots g_L(\uv_i,i) )^t \ $,
and $\albf$ is
the $K \times (L+1)$ matrix of unknown parameters.
Our construction of the nonparametric design matrix specifically 
assumes that the covariates, $\uv_i$, are time independent. If the
covariates are time dependent, the Kronecker product design structure,
$(\gv(\uv_i,i)^t \otimes \Xbf_i)$, should be replaced by 
$\Xbf^i_{G \ p,\ell K +k} = B_k(t_{p,i}) g_{\ell}(\uv_i(t_{p,i}),i)$,
and $\albf$ is replaced by $\alv_G$, the concatenation of the columns
of $\albf$. 

The error structure is assumed to be independent between individuals, but
arbitrary within each individual. 
The $n_i \times n_i$ covariance matrix, $\Sibf_i$, of the
measurement errors may be either given or estimated. Thus our models
include random effects models for the nonparametric part.


To determine the free parameter matrix in our smooth spline growth curve
model, we minimize over ${\albf,\btv}$ the functional:
$$
 \sum_i^N ( \vec{y}_i - \muv_i(\tv_i, \vec{u}_i , \albf,\btv ))^t
\Sibf_i^{-1} ( \vec{y}_i - \muv_i(\tv_i, \vec{u}_i,\albf,\btv ) ) +
\sum_{\ell =0}^L \lambda_{\ell} \int       
|f_{\ell}^{(\gamma)} (t,\alv_{\ell})|^2 dt 
\eqno (4)$$
where $\gamma$ is an integer with $2\le \gamma \le 3$. 
The second term consists of a separate smoothness penalty function for each
of the $f_{\ell}^{(\gamma)}(t).$
The {\it smoothing parameters}, $\lambda_{\ell}$, control the tradeoff
between the goodness-of-fit, 
$\sum_i^N ( \vec{y}_i - \muv_i(\tv_i, \vec{u}_i , \albf,\btv ))^t
\Sibf_i^{-1} ( \vec{y}_i - \muv_i(\tv_i, \vec{u}_i,\albf,\btv ) )$,
and the smoothness of the solution.  
If necessary, the utility functional may be robustified in the standard
ways.
Our analysis generalizes to smoothing surfaces with covariates.
In other words, if time, $t$, is replaced by two nonparametric
variables such as time and age,  $(t,s)$, 
our formulas remain valid if $f_{\ell}{''}(t)^2$ is replaced 
by $|\part_t^2 f_{\ell}|^2 + 2|\part_t\part_s f_{\ell}|^2 +
\part_s^2 |f_{\ell}|^2$.

Each penalty function may be represented parametrically as
$\vec{\alpha}_{\ell} ^t \Sbf  \vec{\alpha}_{\ell} =$
$\int_0^1 f_{\ell}^{(\gamma)}(t)^2 dt$,
where
$
\Sbf_{k,k^{\prime}} \equiv \int_0^1 B_k^{(\gamma)} (t)
B_{k'}^{(\gamma)} (t) dt \ .
$
Differentiating Eq. 4  with respect to $\vec{\alpha}_{\ell}$ yields
$$\sum_{i=1}^N g_{\ell} ( \vec{u}_i,i) \left( \Xbf_i^t \Sibf_i^{-1} \Xbf_i
\alh \vec{g} ( \vec{u}_i,i) -\Xbf_i^t \Sibf_i^{-1} 
(\vec{y}_i - \Hbf_i \btvh ) \right)
+ \lambda_{\ell}  \Sbf \alvh_{\ell} \ = \ 0
\ .\eqno (5a)$$
The corresponding variation of Eq. 4 with respect to $\btv$, the parametric
part of the model, yields:
$$ \sum_{i=1}^N  \Hbf_i^t \Sibf_i^{-1} \Hbf_i
\btvh = \sum_{i=1}^N  \Hbf_i^t \Sibf_i^{-1} 
\left( \vec{y}_i -  \Xbf_i \alh \gv(\vec{u}_i,i) \right)
\ .\eqno (5b)$$
\ni
This is a linear system of $(L+1)K +J$ unknowns. 
The optimal values of the $(L+1)$
smoothing parameters, $\lambda_{\ell}$, are unknown and also may be
determined from the data. 


We now reformulate Eq. 5 as a generalized ridge
regression with $K(L+1)+J$ unknowns and $N_T$  measurements,
where $N_T \equiv \sum_{i=1}^N n_i$.
We concatenate the columns of the $n$ individual measurements into
a $N_T$ vector, $\Yv_c$, and the columns of $\albf$ and $\btv$ into a $K(L+1)+J$
vector $\alv_c:\ \alv_c^t\equiv (\alv_o^t,\alv_2^t...\alv_L^t,\btv^t)$. 
The $N_T \times (K(L+1)+J)$ design matrix consists of the
concatenation of the $N$ matrices $(\gv(\uv_i,i)^t \otimes \Xbf_i, \  \Hbf_i)$,
or for time dependent covariates, $(\Xbf_{G,i}, \  \Hbf_i)$.

The total covariance matrix, $\Sibf_c$, consists of diagonal matrix
entries, $\Sibf_i$. The total penalty function, $\Sbf_c(\lav)$,
consists of diagonal matrix entries $\lambda_{\ell} \Sbf$:
$\Sbf(\lav)_{\ell K + k, \ell'K + k'} =\lambda_{\ell} \Sbf_{k,k'}
\delta_{\ell,\ell'}$ for $0\le \ell \le L$ and zero otherwise.
In this formulation, Eq. 5 is transformed to
$$
\left( \Xbf_c^t \Sibf_c^{-1} \Xbf_c + \Sbf_c(\lav) \right) \alvh_c(\lav)
= \Xbf_c^t \Sibf_c^{-1} \vec{Y}_c 
\ . \eqno (6)$$
\ni
Our interactive spline models are a special class of mixed models.
As such, the covariance structure may be
parametrized with a free parameter vector, $\theta$: 
$\Sibf_c = \Sibf_c(\theta)$, and $\theta$ may be estimated
using a restricted maximum likelihood estimator (Harville (1977)). 
The Bayesian posterior covariance of the discretized system is
$$ 
 Cov \left( \alv_c\alv_c^t \right)
\equiv
\left( \left( \Xbf_c^t\Sibf_c^{-1}\Xbf_c \right)^{-1} 
+\Sbf_c(\lav)^{-} \right)^{-1}
,\eqno (7)$$
where $\Sbf_c(\lav)^{-}$ is the Moore-Penrose generalized inverse of
$ \Sbf_c(\lav)$.

\  \\
\ni
{\bf III. DATA-BASED PARAMETER DETERMINATION}
\bigskip

The optimal values of the smoothing parameters, $\lav$, are unknown, 
and we endeavor to estimate them from the data.
Most data-based estimates determine the
free parameters by minimizing a functional of the data.
Risk-based methods minimize a functional related to the predictive
error, appropriately weighted (Hall and Titterington (1987)). 
Goodness of fit methods minimize the residual squared error
weighted by a measure of the number of effective degrees of freedom.

 

We now present  several 
common goodness of fit functionals, and then  a class of risk-based
functionals. These functionals do not
require the growth curve structure which we discussed previously.
Instead, we require only that the covariance structure, $\Sibf_c$,
is known up to an arbitrary scalar: $\Sibf_c = \sigma^2 \Sibfh$.
We also assume that the penalty matrix, $\Sbf_c(\vec{\lambda})$
satisfies
$ \Sbf_c(\vec{\lambda}) = \sum_{\ell} \lambda_{\ell} \Sbf_{\ell}$.
We redefine $\lav$: $\lav^{new}\equiv \lav^{old}\sigma^2$.
We define $\Cbf\equiv \Xbf_c^t \Sibfh_c^{-1} \Xbf_c$, and 
the matrices, $\Gbf(\lav)$, and
the influence matrix, $\Abf_c(\lav)$: 
$$ \
\Gbf(\lav)\equiv
\left( \Xbf_c^t \Sibfh_c^{-1} \Xbf_c + \Sbf_c(\vec{\lambda}) \right)^{-1} 
\ , \
\Abf_c(\lav)\equiv
\Xbf_c \Gbf(\vec{\lambda})
\Xbf_c^t \Sibfh_c^{-1}
.\eqno (8)$$
\ni 
The influence matrix is not a projection due to the penalty matrix,
$\Sbf_c(\vec{\lambda})$: $\Ab_c(\lav) >\Ab_c(\lav) \Ab_c(\lav)$, and 
therefore the concept of effective degrees of freedom is tenuous. 

Most of the goodness of fit data-based functionals (when $\sigma^2$
is unknown) are of the form:
$$
V(\lav) \equiv
\frac{\parallel (\vec{Y}_c -\Abf_c (\lav)\Yv_c)^t \Sibfh^{-1}
(\vec{Y}_c -\Abf_c (\lav)\Yv_c) \parallel}
{N_T M(\Abf_c(\lav))} 
,\eqno (9)
$$
where $M$ is a real valued function on $N_T\times N_T$ matrices.
Furthermore, these functionals usually satisfy
${M(\Abf_c(\lav))} \sim 1.0 - 2Trace{(\Abf_c(\lav))}/N_T$ when
$N_T>> Trace{(\Abf_c(\lav))}$ 
(Hardle, Hall, and Marron (1988)).
Generalized cross-validation (G.C.V.)
(Craven and Wahba (1979)) is the most widespread 
data-based functional of the
form given by Eq. 9,  with 
${M(\Abf_c(\lav))} \equiv |1.0 - Trace{(\Abf_c(\lav))}/N_T|^2$.
Another goodness of fit functional is the corrected Akaike information
criteria  (Hurvich and Tsai (1989)), which is based on
the information content.
G.C.V. automatically includes the effects of model error including 
discretization error in its estimate of the optimal $\lav$.
In contrast, estimates of the 
expected error often neglect model error including discretization error.
 
The total expected error in $\alv_c(\lav)$ from the sampling perspective, 
assuming the discrete model is correct, is
\ \\
$ E \left[ (\alvh_c(\lav)-\alv_c) (\alvh_c(\lav)-\alv_c)^t \right]
\ = $
$$
\ \sigma^2
\Gbf(\lav)
\Cbf \Gbf(\lav)
\ + \ 
\Gbf(\lav)\Sbf_c(\lav)\alv_c\alv_c^t \Sbf_c(\lav) \Gbf(\lav)
.\eqno (10a)
$$
\ni
$\sigma^2\Gbf(\lav)\Cbf  \Gbf(\lav)$ is the variance,
and 
$\Gbf(\lav)\Sbf_c(\lav)\alv_c\alv_c^t \Sbf_c(\lav) \Gbf(\lav)$
is the bias error. For any positive
semidefinite matrix, $\Qbf$, we can select $\lav$ by minimizing
\ \\
$R(\lav,\Qbf)= Trace(\Qbf E \left[ (\alvh_c(\lav)-\alv_c) 
(\alvh_c(\lav)-\alv_c)^t \right])  =$ 
$$Trace(\Qbf  \left[ \sigma^2  
\Gbf(\lav)\Cbf  \Gbf(\lav)
\ + \ 
\Gbf(\lav)\Sbf_c(\lav)\alv_c\alv_c^t \Sbf_c(\lav) \Gbf(\lav)
 \right]) \ , \eqno (10b)$$
with respect to $\lav$. When $\Qbf\equiv \Xbf^t \Sibf^{-1}\Xbf$,
the risk estimate corresponds to minimizing the predictive error.  
Choosing $\Qbf= \Sbf_c$ corresponds to minimizing the 
average expected square error in estimating the $\gamma$th radial derivative.
From the asymptotic results of Cox (1984), we expect that derivative 
estimation will require larger values of $\lambda$ than estimating the
unknown profiles. 
Equations 9 and 10 are generalizations
of previously known functionals  (Craven and Wahba (1979))
to an arbitrary covariance matrix, $\Sibf_c$, and similar equations
are given in Diggle and Hutchinson (1989). 
When $\lav$ is selected to minimize the risk, we have:
$$
{1\over 2}{\part R(\lav,\Qbf)\over \part \lambda_{\ell}}= 0 = 
Trace\left[    \Qbf  \Gbf(\lav) \Sbf_{\ell} \Gbf(\lav)
\left( \Cbf\alv_c\alv_c^t \Sbf_c(\lav)- \sigma^2\Cbf \right) \Gbf(\lav) 
 \right] , \eqno (11)$$
\textheight  8.17 in
\renewcommand{\baselinestretch}{1.45}
\ni
When $\sigma^2$ is known, but $\alv_c$ is unknown, we have the  following
estimate for  the  minimum of the expected loss:
\ \\
${1\over 2}{\part \Rh(\lav,\Cbf)\over \part \lambda_{\ell}}= 0 = $ 
$$
\vec{Y}_c^t\Sibfh^{-1}\Xbf_c \Gbf(\lav) \Sbf_c(\lav)  \Gbf(\lav) \Sbf_{\ell}
\Gbf(\lav)\Xbf_c^t \Sibfh^{-1} \vec{Y}_c - \sigma^2
Tr\left[ \Gbf(\lav) \Sbf_{\ell} \Gbf(\lav) \Cbf
 \right] \ , \eqno (12)$$
where we have restricted to $\Qbf = \Cbf$.
Equation 11 or 12 constitute a set of $L + 1$ equations to determine
the  optimal values of $\{ \lambda_{\ell} \} $.
When $\Sibf$ is known and the errors are uncorrelated, the
risk-based estimate, $\Rh(\lav,\Cbf)$
 was proposed in Craven and Wahba (1979).
We prefer the differential formulation of Eq. 12,
${1\over 2}{\part \Rh(\lav,\Cbf)\over \part \lambda_{\ell}}= 0 $,
because Eq.  12  shows explicitly that minimizing $\Rh$ is a weighted
goodness of fit estimator.  

In practice, $\sigma^2$ is often unknown, and
can be estimated from the data using 
$$
\sgh^2 =
\frac{\parallel (\vec{Y}_c -\Abfh_c (0)\Yv_c)^t \Sibfh^{-1}
(\vec{Y}_c -\Abfh_c (0)\Yv_c) \parallel}
{[N_T-{\rm tr}(\Abfh_c(0))]} ,
\eqno (13)
$$
where $\Abfh_c (0) = \Xbf_c  \Cbf \Xbf_c^t \Sibfh^{-1}$.
The variance in the estimate of $\sigma^2$ of Eq. 13 is  inversely
proportional to $[N_T-{\rm tr}(\Abfh_c(0))]$, and therefore increases
as the number of spline basis functions grow. 
We can choose the number of basis functions to minimize the tradeoff of
variance in $\sgh^2$ to bias in the discretization error. 

The number and choice of 
the basis functions, $h_j(t,\uv,i)$, and 
$g_{\ell}(\uv)$, may also be determined by minimizing 
the data-based functional.
However, as the dimension of
the multivariate minimization increases,
these data-based functionals may
have a number of relative minima. 
Furthermore, the minimum of the function is often shallow,   
and the estimated value of $\lav$ may converge slowly
to its optimal value as $N$ tends to infinity. 
The convergence to the optimal value is slow when the utility
function, $V(\lambda)$, is an insensitive function of the smoothing.
On the other hand, in these cases, 
the risk/goodness of fit is not dramatically worsened  
by the use of a suboptimal value of $\lav$.

\ \\
\ni
{\bf IV.  NUMERICAL IMPLEMENTATION}
\medskip

In our numerical implementation, we concatenate the rows of the
$K \times (L+1)$ matrix $\alpha$ instead of the columns. The corresponding
$n_i \times K(L+1)$ design matrix for the $i$th individual  satisfies
$X_{{j,(L+1)(k-1) + \ell+1}}^i = B_k (r_j^i )g_{\ell} (u_i )$
where $j = 1 \ldots n_i$.
The advantage of this reordering of the unknown coefficients is that the
resulting $\Xbf^t \Xbf$ matrix has a band structure.

The empirical estimate of the minimizing risk, Eq. 12 can be rewritten as
$$ \lambda_{\ell} = {\sigma^2
Tr\left[ \Gbf(\lav) \Sbf_{\ell} \Gbf(\lav) \Cbf \right] -
\sum_{\ell' \ne \ell} \lambda_{\ell'}
\alvh{}^t  \Sbf_{\ell'}  \Gbf(\lav) \Sbf_{\ell} \alvh
\over \alvh{}^t  \Sbf_{\ell}  \Gbf(\lav) \Sbf_{\ell} \alvh}
\ , \eqno (14a)$$
where $\alvh$ is given in Eq. 6. 
We iteratively evaluate Eq. 14a for each $\lambda_{\ell}$ to minimize Eq. 12.

Two simplifications of the $\lambda$ estimate are possible. First,
when $\Gbf(\lav)$ is approximately block diagonal, we can neglect
the second term in the numerator of Eq. 14a, and the simplified equation is 
$$ \lambda_{\ell} = {\sigma^2
Tr\left[ \Gbf(\lav) \Sbf_{\ell} \Gbf(\lav) \Cbf \right] 
\over \alvh{}^t \Sbf_{\ell}  \Gbf(\lav) \Sbf_{\ell} \alvh}
\ . \eqno (14b)$$
In practice, the variance in the estimated mean profile, $f_0 (t)$, is
usually much smaller than the variance in the estimates of $f_1 (t)
\ldots f_{\ell} (t)$. Thus if all $\lambda_{\ell}$ are equal, the
smoothing tends to be too little for $f_0 (t)$ or too much for $f_1
(t) \ldots f_{\ell} (t)$. Thus a second simplification is to
reduce the dimensionality of the minimization
by imposing the model restriction: 
$\lambda_1 \equiv \lambda_2 \ldots \lambda_{\ell}$ and
$\lambda_o \ne \lambda_1$.

\ \\
\ni
{\bf V.  EXAMPLE}
\medskip

We consider a 40 profile dataset from the Tokamak Fusion Test Reactor 
(T.F.T.R.) at the Princeton Plasma Physics Laboratory
(Hiroe et al. (1988). 
Each profile consists of approximately 61 temperature measurements
at different spatial locations. In contrast,the data from 
our earlier study of the A.S.D.E.X. tokamak had only 15 spatial locations.
Thus the A.S.D.E.X data  was well modeled with a spline with only three 
knots while the T.F.T.R. data is better suited to a smoothing spline
ten or more knots.

In the middle of the plasma, the error bars are proportional to the 
temperature, while the errors at boundary are roughly constant.
Thus we use the logarithm of the temperature as the dependent variable
and increase $\Sigma_{k,k}$  towards the boundary.
We estimate $\sigma^2$ by fitting each profile separately to a one dimensional
model.  
We use $\gamma =3$ in the penalty function.
To determine $\lambda_0$ and $\lambda_1$, we iterate Eqs. 6 \& 14b. 

Figure 1 plots the raw data and the fitted profile for the two profiles
with the largest and smallest value of $q$ for the fourteen knot fit.
Figure 2 gives the corresponding fit with 46 knots. The similarity of the
two fits supports our assertion that smoothing spline fits are often
only weakly dependent on the choice of knots. Similarly, the fits
are also insensitive to the values of $\lambda_0$ and $\lambda_1$.
The error bars in Figs. 1 \& 2 are given by the ``plug-in'' approximation, 
i.e. we  substitute $\alvh$ into Eq. 10a
to estimate the local value of the expected square error for the
profiles fits. The expected error increases near the boundaries, partly because
the measurement variance increases near the boundary.  
To reduce the boundary effect, 
we have increased the spacing between knots near the
boundary. The vertical lines on Figs. 1 \& 2 give the knot locations.

We have fit the profiles with up to four covariates, plasma $q$, 
average particle density, average magnetic field, and electrical voltage.
The additional covariates do not produce any noticable change in the 
predicted profiles.

The slight misfits near $r = 2$ for the low $q$ profile and
near $r = 2.5$ for the high $q$ profile appear to be due to random
variation in the profile shape rather than systematic errors in the 
additive spline model.
For a given profile, the residual errors tend to be uniformly positive or
negative, indicating that the shape of the temperature profile is better
determined than the magnitude of the average temperature.
We are currently implementing the random intercept covariance model:
$\Sigma_{j,k} = \sigma^2 \delta_{j,k} + \sigma_o^2$. 
The model of Eq. 2b is the corresponding fixed effects model for
the intercept. 
For a more detailed discussion of the experimental findings, we
refer the reader to (Imre and Riedel (1993)).

\ \\
\ni
{\bf VI.  SUMMARY}
\medskip

Growth curves  
characteristically have much temporal structure and resolution and  
relatively poorly resolved covariate structure. Thus we apply
nonparametric smoothing splines in the temporal representation and
simple, parametric representations in the covariate directions.
This class of models does not require multiple time measurements, 
but the class is well suited to this structure.
In other situations, there may be only enough data in the temporal
direction for parametric representations or sufficient data in the
covariate directions for a nonparametric representation.

We close by noting that Eq. 12 is the general matrix weighting of the residual
error that estimates the value of $\lambda$ which minimizes the
expected loss. Under the assumption that $\Cbf$ and $\Sbf$ 
are equal up to a scalar multiple, Hall \& Titterington (1987) have shown that
estimating minimizer of the expected loss is equivalent to a specific 
goodness of fit estimator. Equation 12 generalizes the Hall \& Titterington
result to the case when $\Cbf \ne c \Sbf$. 

\ \\
\ni
{\bf APPENDIX: UNIFORM TEMPORAL DESIGN}
\medskip

When the temporal design is uniform,
i.e. $\Xbf_1 = \Xbf_2 \ldots = \Xbf_N$, Eq. 5a may be recast in 
multivariate form. For simplicity, we assume $\btv \equiv 0$ in this appendix.
The $N$ observed profiles, each consisting of  the same $n$ time points,
can be represented by a $ n \times N$ 
data matrix, $\Ybf$. The  $(L+1) \times N$ covariate data matrix, $\Ubf$, 
has columns $\gv(\uv_1)$ through $\gv(\uv_N)$. 
In multivariate notation, the growth curve model is
$${\Ybf = \Xbf \albf \Ubf + \Ebf,
}\eqno (15)$$
where 
$\Ebf$ is the $n \times N$ matrix of random errors.
The first variation of the the utility functional (Eq. 5a)
can be rewritten as 
$$  \left( \Xbf^t \Sibf^{-1} \Xbf
\albf \Ubf \Ubf^t 
+ \Sbf \albf \Labf \right) =
 \Xbf^t \Sibf^{-1} \Ybf \Ubf^t
,\eqno (16)$$
where $\Labf$ is a $(L+1)\times (L+1) $ diagonal matrix whose
elements are $\lambda_{\ell}$.
Unfortunately, the tensor product formulation does not allow 
a solution based on the separation of variables. Instead, the
full $(L+1)K \times (L+1)K$ system must be solved. 

A separable smoothing spline model may be constructed as follows.
We decompose $\Ubf \Ubf^t$ into $\Obf \Dbf \Obf^t$ where
$\Obf$ is orthonormal and $\Dbf$ is diagonal. We then replace
the growth curve model of Eqs. 2 \& 3 with the equivalent model
with $ \albf^{new}= \albf^{old} \Obf$ and 
$\vec{g} ( \vec{u}_i,i)^{new}=\Obf^t \vec{g}(\vec{u}_i,i)^{old}$.
This is equivalent to replacing the old penalty function for
the functions $f_{\ell}(t)$ with a matrix valued penalty function.
The separable system splits the growth curve problem into
$L+1$ independent one dimensional problems. For the separable
problem, the  one dimensional convergence of 
Cox (1984) results may be extended trivially.

\ \\ \ni
{ ACKNOWLEDGMENTS}

\medskip
KSR's understanding of growth curves has benefited
from years of collaboration with O. Kardaun and P. McCarthy.
We thank C. Hurvich 
for discussions on data-based selection criteria.
D. Stevens's help in the initial implementation of the spline fit is 
gratefully acknowledged.
The T.F.T.R. data was collected by the T.F.T.R. experimental
team at Princeton University. We received the data  from 
the U.S. D.O.E.-I.T.E.R. magnetic fusion energy database. 
We thank both the T.F.T.R. and M.F.E. database groups, in particular
Dr. W.H. Miner Jr. 
The valuable comments of the referee are appreciated.
This work was supported by the U.S. Department of Energy. 

\np
BIBLIOGRAPHY 
\begin{enumerate}{}

\item{Agarwal, G. and Studden, W. (1980).
Asymptotic mean square error using least squares and bias minimizing splines.
{\it Annals of Statistics} {\bf 8}, 1307-1325. }

\item{Breiman, L. (1991).
The $\Pi$ method for estimating multivariate functions from noisy data.
{\it Technometrics} {\bf 33}, 125-160}


\item{Cox, D.D. (1984). 
{Multivariate smoothing splines functions.}
{\it SIAM J. Numer. Anal.} {\bf 21}. 789-813. }

\item{
Craven, P. and Wahba, G. (1979).
Smoothing noisy data with spline functions: estimating the correct
  degree of smoothing by the method of generalized cross-validation.
{\it Numer. Math.} {\bf 31}, 377-403.}

\item{
deBoor, C. (1978).
{\it A Practical Guide to Splines}. Springer-Verlag, New York.}

\item{Diggle, P.J. and Hutchinson, M.F. (1989). 
{Spline smoothing  with autocorrelated errors.}
{\it Australian J. Stat.} {\bf 31}, 166-182. }

\item{
Eubank, R. (1988). 
{\it Smoothing  splines and nonparametric regression.}
Marcel Dekkar, N.Y., Basel.}


\item{
Friedman, J.H. (1991). 
{Multivariate adaptive regression splines (with discussion).}
{\it Annals of Statistics} {\bf 19}, 1-143. }

\item{
Friedman, J.H., and Stuetzle, W. (1981). 
{Projection pursuit regression.}
{\it J.A.S.A.} {\bf 76}, 817-823. }


\item{
Gasser, T., Muller, H.-G., Kohler, W. et al. (1984). 
{Nonparametric regression analysis of growth curves.}
{\it Annals of Statistics} {\bf 12}, 210-229. }

\item{ Geisser, S. (1980). Growth Curve Analysis.
{\it Handbook of Statistics}, P.R. Krishnaih, ed.,
Vol.1,  89-115, North Holland Publishing Co.,  Amsterdam. 
}
 
\item 
{Grizzle, J.E. and Allen, D.M (1969). 
Analysis of growth and dose response curves. 
{\it Biometrics} {\bf 25},  357-381.}

\item{Gu, C. and Wahba, G. (1991).
Smoothing spline ANOVA with componentwise Bayesian ``confidence
intervals''.
Technical Report 881, Department of Statistics, University of
  Wisconsin, Madison.}



\item
{ Hall, P., and Titterington, D. 
(1987). {Common structure of techniques for choosing smoothing 
parameters in regression problems.}
{\it J. Roy. Stat. Soc. Ser. B} {\bf 49}, 184-198. }


\item
{ Hardle, W., Hall, P., and Marron, S. 
(1988). {How far are automatically chosen smoothing 
parameters from their optimum?}
{\it J.A.S.A.} {\bf 83}, 86-95. }


\item{Harville, D.A. (1977).
Maximum likelihood approaches to variance component
estimation and related problems.
{\it J.A.S.A.} 
{\bf 72},  320-340. }

\item
{Hastie, T.  and Tibshirani, R. (1990).
{\it Generalized Additive Models.}
Chapman Hall, London.}

\item{
Heckman, N.   (1986). 
{Spline smoothing in partly linear models.}
{\it J. Roy. Stat. Soc. Ser. B} {\bf 48}, 244-248. }

\item{Hiroe, A., et al., (1988).
Scale length study in T.F.T.R.
{\it  Princeton Plasma Physics Laboratory Report \# 2576.}
}

\item 
{Hurvich, C.M. and Tsai, C.-L. (1989). 
Regression and time series model selection
in small samples.
{\it Biometrika} {\bf 76},  297-307.}

\item
{ Imre, K., and Riedel, K.S.  (1993). In progress. }  

\item 
{Kardaun, O., McCarthy, P.J., Lackner, K., 
Riedel K.S., and Gruber, O.
(1988).
{ A statistical approach to profile invariance}.
{\it  Theory of Fusion Plasmas (Va\-ren\-na 1987 Proc.),
Societ\`a Italiana di Fisica, Bologna}, 435-444. }





\item{
Lee, J.C., and Geisser, S. (1972). 
{ Growth curve prediction.} 
{\it Sankhya Ser. A} {\bf 34}, 393-412. }


\item 
{McCarthy, P.J., Riedel, K.S., Kardaun, O., et al.
(1991).
{ Scalings and plasma profile parameterisation of ASDEX high density
Ohmic discharges}.
{\it Nuclear Fusion} {\bf 31}, 1595-1633 .}

\item 
{Muller, H. (1988). {\it Nonparametric Regression Analysis of
Longitudinal Data}, 
Lecture Notes in Statistics
Vol. 46, Springer, Heidelberg. }

\item 
{Potthoff, R.F. and Roy, S.N. (1964).
A generalised multivariate analysis of variance model useful 
especially for growth curve problems. 
{\it Biometrika} {\bf 51}, 313-326.} 

 
  

\item 
{Rao, C.R. (1975).  
Simultaneous estimation of parameters in different linear models
and applications to biometric problems.
{\it Biometrics} {\bf 31}, 545-554.}

\item 
{Raz, J. (1989). 
Analysis of repeated measurements using nonparametric smoothers 
and randomisation tests.
{\it Biometrics} {\bf 45}, 851-871.}

\item{
Rice, J.B. (1986). 
{Convergence rates for partially splined models.}
{\it Statist. Prob. Lett.} {\bf 4}, 203-208. }



\item{Riedel, K.S. (1992).
Smoothing spline growth curves with covariates.
{\it Courant Institute of Mathematical Sciences Report MF-123, 
New York University.}}


\item{
Silverman, B. (1985). 
{Some aspects of the spline smoothing approach to nonparametric
regression curve fitting.}
{\it J. Roy. Stat. Soc. Ser. B} {\bf 46}, 1-52. }

\item{
Speckman, P.E. (1988). 
{Regression analysis for partially linear models.}
{\it J. Roy. Stat. Soc. Ser. B} {\bf 50}, 413-436.}

 
\item 
{Strenio, J.F., Weisberg, H.I., and Bryk, A.S. (1983). 
{Empirical Bayesian estimation of individual growth curve
parameters and their relationship to covariates.}
{\it Biometrics} {\bf 39}, 
71.}



\item{
Wahba, G. (1990).
{\it Spline Models for Observational Data},
S.I.A.M., Philadelphia.}


\item 
{Wegman, E. and Wright, I. (1983).  
Splines in statistics.  
{\it J.A.S.A.} {\bf 78}, 351-366.}


\end{enumerate}{}

\end{document}